# Cluster transfer matrix method for the single electron box

S.G. Chung

Department of Physics, Western Michigan University, Kalamazoo, MI 49008-5151

Abstract

With the newly developed cluster transfer matrix method, we calculate the average electron number n vs $n_x$ (the polarization charge) for varying junction conductance and its first derivative at $n_x=0$ for finite temperatures, and demonstrate that the new method is as powerful as the Monte Carlo and renormalization group methods.

The single electron box (SEB)[1] is the most elementary device which exhibits the Coulomb blockade phenomenon.[2,3] It consists of a metallic island connected to an outside lead by a tunnel junction and coupled capacitively to a gate voltage, and described by the Hamiltonian[4]

$$H = \frac{1}{2}(n-n_x)^2 + \sum_{ab} t_{ab} c_a^+ c_b + h.c. + H_0 \qquad (1)$$

where $n_x$ is a continuous polarization charge induced by the gate, n is the electron number operator in the box, $t_{ab}$ is the transfer integral for electron transfer between the box(index a) and the lead(index b), and $H_0$ describes noninteracting reservoir fermions. After integrating out fermion degrees of freedom, the partition function can be written as a path integral over a phase variable φ which is conjugate to the electron number n,[4,5]

$$Z(n_x) = \sum_{m=-\infty}^{\infty} e^{2\pi i m n_x} \int_{-\infty}^{\infty} d\phi_0 \int_{\phi_0}^{\phi_0 + 2\pi i m} \mathcal{D}\varphi$$

$$\exp\left\{ -\int_0^\beta d\tau \frac{1}{2}\dot\phi^2 - \frac{2g}{\pi^2} \int_0^\beta \int_0^\beta d\tau d\tau' \frac{\sin^2\left(\frac{\phi(\tau)-\phi(\tau')}{2}\right)}{(\tau-\tau')^2} \right\} \qquad (2)$$

where $g = R_Q/R_t$, with the quantum resistance $R_Q = \frac{h}{4e^2}$ and a junction resistance $R_t$, is the dimensionless tunneling conductance. β is inverse of the temperature T (we put the Boltzmann constant=1). The problem is the competition between the charging effect, the first term, and temperature T and the tunneling effect, the second term in the exponent in (2). At T = 0 and without the tunneling effect, n vs $n_x$ is a perfect staircase. It is also clear that for $T > e^2/2C$ or large enough g, the Coulomb blockade is absent. The question is then how in general T, g and $n_x$ affect n. Theoretical studies so far have been based on perturbation theories, perturbative renormalization group, the instanton method, scaling theories and Monte Carlo simulations.[5-11]

In a recent paper[12], we proposed a new method, the cluster transfer matrix (CTM) method, for the SEB. The method is for arbitrary polarization charge $n_x$, temperature T and junction conductance g. In Ref. 12 we concentrated on the $n_x = T = 0$ case. We here calculate the average electron number n vs $n_x$ for varying g and its first derivative at $n_x=0$ for finite T, and demonstrate that the new method is as powerful as the Monte Carlo (MC) and renormalization group (RG) methods.

We adopt the same notations as in Ref. 12. We will not repeat the formulation here, but briefly discuss the basic ideas and point out some corrections to the equations in Ref. 12. The interval (0, β) is first divided into N islands and each island contains M discrete points So the lattice constant $\Delta = \beta/MN$. We rewrite the integrand in (2) as a product of two terms. One describes the nearest neighbor interaction between islands and the other the remnant. The former is written as

$$\prod_{i=1}^{N} K(\vec{\phi}_i, \vec{\phi}_{i+1}) \tag{3}$$

where $\vec{\phi}_i$, i = 1, 2, .., N denote M dimensional vectors describing the dynamics of M phase variables in an island. We then consider a CTM eigenvalue problem for the operator K, which is essentially a quantum M body problem with a free center of mass motion. The latter fact indicates a generalized Bloch-Floquet theorem. The CTM quantum number is thus decomposed into k for the center of mass motion and ℓ for the relative motion. After some manipulations and repeatedly using the CTM equation, we arrive at

$$Z(n_x) = R \sum_{\ell,m} \lambda_{\ell,n_x+m}^{N} \tag{4}$$

where λ is a CTM eigenvalue and a summation about m is over integers. The remnant R is calculated by a cumulant expansion. With the cluster size M, both the perturbative treatment of R and its cumulant evaluation, thereby neglecting higher order correlations, will become increasingly correct.

The procedure for calculating the correlation function in the remnant R is the same as in Ref. 13. The two expressions, Eq (13) and (14) in Ref. 12, however, should be corrected as

$$\langle \cos\phi(\tau)\cos\phi(\tau') \rangle = \frac{1}{\sum_{m,n} \lambda^N_{m,n_x+n}} \sum_{m,m',k',n} \lambda^{N-(q-p)}_{m,n_x+n} \cdot \lambda^{q-p}_{m',k'} \cdot W_c \quad (5)$$

where we have assumed that $\tau$ is the i-th lattice point in the p-th island and $\tau'$ j-th lattice point in the q-th island, and

$$W_c(m,m') = \sum_{n'} C_{n',m'} \langle \overline{\psi}_{n',k'} | \cos\phi(\tau) | \psi_{m,n_x+n} \rangle \cdot \sum_{n''} C_{n''m} \langle \overline{\psi}_{n'',n_x+n} | \cos\phi(\tau') | \psi_{m',k'} \rangle \quad (6)$$

The $\langle \sin\phi(\tau)\sin\phi(\tau') \rangle$ can be calculated similarly and found equal to $\langle \cos\phi(\tau)\cos\phi(\tau') \rangle$. In the last equation, $\Psi$ is the CTM eigenvector and $\overline{\psi}$ is its Schmidt orthogonalized one. Once the partition function and hence the free energy f is obtained, then the average electron number in the box is calculated by

$$\langle n \rangle = n_x - \frac{\partial f}{\partial n_x} \quad (7)$$

For $n_x = 0$, a quantity which directly measures the degree of Coulomb blockade is the slope $\chi = \partial\langle n \rangle / \partial n_x$. $\chi = 0$ means perfect blockade, while $\chi = 1$ its absence.

There are two important parameters in the theory, the lattice constant $\Delta$ and the cluster size

M. The larger the $\Delta$, the faster the convergence of physical quantities such as $\chi$ with increasing M. The theory contains in itself a means to find maximum efficient $\Delta$ when other parameters are specified. That is to calculate, say $\chi$, for a $\Delta$ by some M and for $\Delta' = \Delta M/M'$ by a larger M'. If the obtained $\chi$ and $\chi'$ agree within an allowable relative error, we are not missing important degrees of freedom by choosing that $\Delta$.

In Ref. 12 we calculated $\chi$ for the case T = 0 and $n_x$ = 0. Fig 1 is a replot of Fig 2 in Ref. 12 for g≤6 where converged and hence exact result is obtained together with some recent results due to other methods. Our result agrees with the perturbation theory for g≤3.5, the RG result for g≤5, and the MC=instanton result near g=6. At present, to our knowledge, no reliable methods are available for finite temperatures. The CTM can handle finite temperatures equally well. Fig 2 shows the M = 4 result for $\chi$ vs g with $n_x$ = 0 and T = 0.003(square), 0.1(circle) and 0.2(triangle). It is noted that our M = 2 cluster calculation applied to the g = 0 case perfectly reproduces the exact result for arbitrary temperature. Finally, Fig 3 shows the average electron number n vs $n_x$ at T=0 and g=1(squares), 3(circles) and 5(triangles). Except for the $n_x$=0.5 vicinity for g=5 where our result is slightly (1%) too large compared to the expected straight line n=$n_x$, the result is in excellent agreement with the recent RG theory.[7]

In conclusion, we have calculated some important quantities for the single election box by the cluster transfer matrix method. The most notable point of the method is that it is systematic and exact as far as it converges with the cluster size. In fact, convergence in M means that a part of the remnant calculated perturbatively with cumulant expansion and treated exactly through the CTM equation agree, indicating the exactness of the CTM calculation of the partition function. Moreover it contains in itself a means to determine the maximum efficient lattice constant, which drastically accelerates the convergence, enabling us to reach a fairly wide parameter space. Furthermore the method can handle any gate voltage, temperature and junction conductance equally well.

I thank Shingo Katsumoto for stimulating discussions. This work was partially supported by NSF under DMR 980009N and utilized the Silicon Graphics Origin2000 at the National Center for Supercomputing Applications, University of Illinois at Urbana-Champaign.

**References**

1. P. Lafarge, H. Pothier, E.R. Williams, D. Esteve, C. Urbina, and M.H. Devoret, Z. Phys. B**85**, 327 (1991).

2. D.V. Averin and K.K. Likharev, in *Mesoscopic Phenomena in Solids*, edited by B.L. Altshuler, P.A. Lee and R.A. Webb (Elsevier, Amsterdam, 1991), p. 173.

3. M.H. Devoret and H. Grabert in *Single Charge Tunneling*, edited by H. Grabert and M.H. Devoret, NATO Advanced Study Institutes, Ser. B, Vol. 294 (Plenum Press, New York, 1992).

4. G. Schön and A.D. Zaikin, Phys. Rep. **198**, 237 (1990).

5. W. Hofstetter and W. Zwerger, Phys. Rev. Lett. **78**, 3737 (1997).

6. G. Göppert and H. Grabert, cond-mat/9802248.

7. J. König and H. Schoeller, cond-mat/9807103.

8. S.V. Panyukov and A.D. Zaikin, Phys. Rev. Lett. **67**, 3168 (1991).

9. G. Falci, G. Schön and G.T. Zimanyi, Phys. Rev. Lett. **74**, 3257 (1995).

10. X. Wang and H. Grabert, Phys. Rev. B**53**, 12621 (1996); X. Wang, R. Egger and H. Grabert, Europhys. Lett. **38**, 545(1997).

11. C.P. Herrero, G. Schön and A.D. Zaikin, cond-mat/9807112.

12 S.G. Chung, J. Korean Phys. Soc. **33**, S25(1998).

13. S.G. Chung, D.R.A. Johlen and J. Kastrup, Phys. Rev. B**48**, 5049 (1993); S.G. Chung, J. Phys. Condens. Matter **9**, L219 (1997).

**Figure Caption**

Fig.1   1-$\chi$ vs g up to g=6, solid circle.  Dash-dot line: perturbation theory of order $g^3$(Ref 6).  Solid line: renormalization group method(Ref 7).  Dash-dot-dot line:  instanton method (Ref 8).  The open triangles are MC data from Ref 10, whereas the open squares are MC data from Ref 11.

Fig. 2   The slope $\chi$ vs g with $n_x = 0$ and M = 4 for T = 0.003(square), 0.1(circle) and 0.2(triangle).

Fig. 3   n vs $n_x$ at T=0 and for g=1(squares), 3(circles), and 5(triangles).

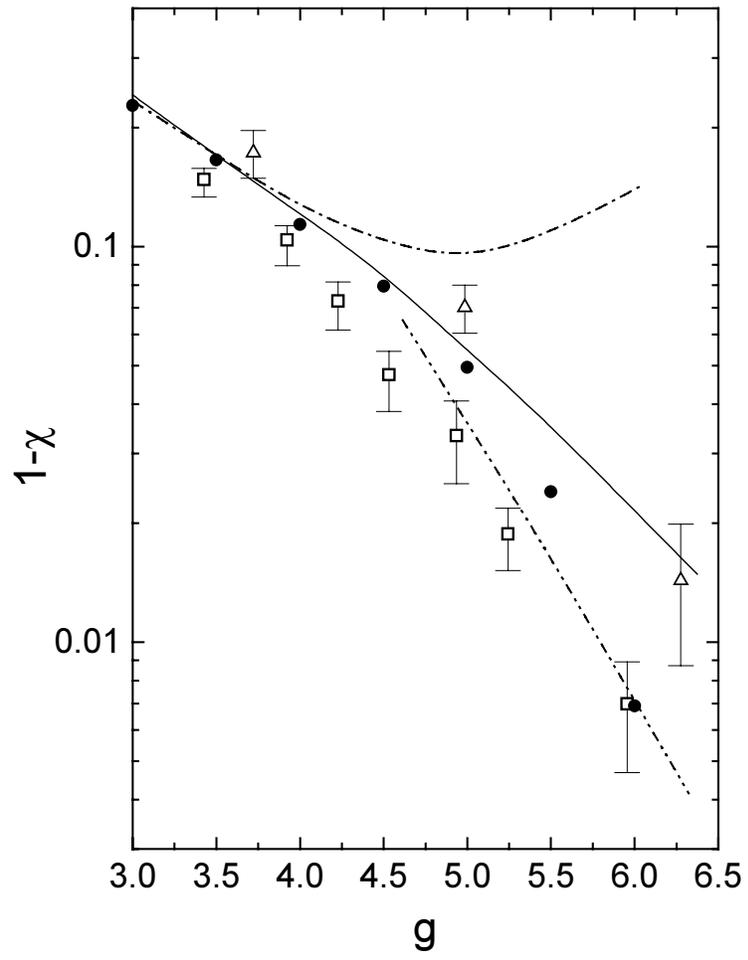

Fig 1

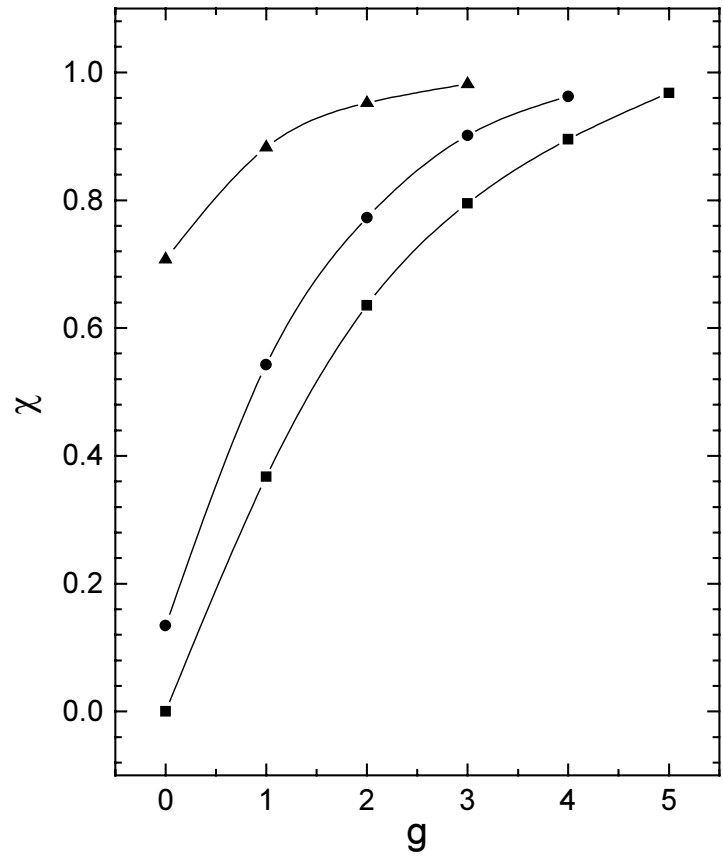

Fig 2

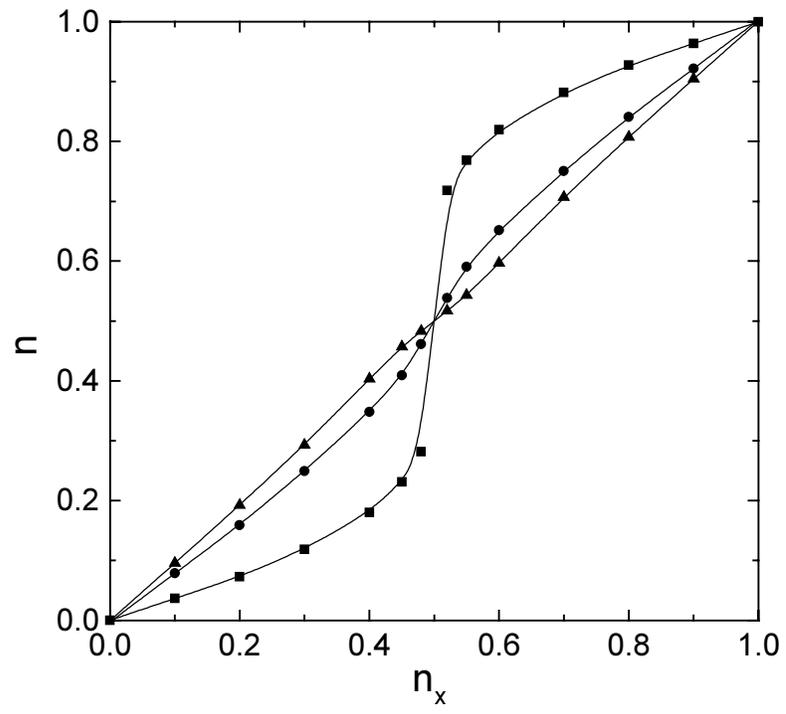

Fig 3